\documentclass[runningheads]{llncs}

\usepackage[T1]{fontenc}        
\usepackage[utf8]{inputenc}     

\usepackage{amsmath,amssymb}    
\usepackage{booktabs}           

\usepackage{graphicx}           
\graphicspath{{figures/}}       

\usepackage{subcaption}         
\usepackage{float}              
\usepackage{url}                

\usepackage{newtxtext,newtxmath}  

\begin{document}
\title{From Cooperation to Hierarchy: A Study of Dynamics of Hierarchy Emergence in a Multi-Agent System}

\titlerunning{From Cooperation to Hierarchy}
%
%
%
\author{Shanshan Mao\inst{1} \and
Peter Tino\inst{1}}

\authorrunning{S. Mao and P. Tino}

\institute{School of Computer Science, University of Birmingham, Birmingham, UK\\
\email{sxm1550@student.bham.ac.uk, p.tino@bham.ac.uk}}

%
\maketitle
\begin{abstract}
A central premise in evolutionary biology is that individual variation can generate information asymmetries that facilitate the emergence of hierarchical organisation.
To examine this process, we develop an agent-based model (ABM) to identify the minimal conditions under which hierarchy arises in dynamic multi-agent systems, focusing on the roles of initial heterogeneity and mutation amplitude across generations.
Hierarchical organisation is quantified using the Trophic Incoherence (TI) metric, which captures directional asymmetries in interaction networks.
Our results show that even small individual differences can be amplified through repeated local interactions involving reproduction, competition, and cooperation, but that hierarchical order is markedly more sensitive to mutation amplitude than to initial heterogeneity.
Across repeated trials, stable hierarchies reliably emerge only when mutation amplitude is sufficiently high, while initial heterogeneity primarily affects early formation rather than long-term persistence.
Overall, these findings demonstrate how simple interaction rules can give rise to both the emergence and persistence of hierarchical organisation, providing a quantitative account of how structured inequality can develop from initially homogeneous populations.

\keywords{hierarchy emergence \and trophic incoherence \and multi-agent systems}

\end{abstract}
\section{Introduction and Background}

From cooperative hunting in African social animals \cite{detrain2008collective,couzin2009collective}
to the stratification of modern human societies \cite{lane2006hierarchy},
hierarchical structures—often implicit \cite{gagnon2011high}—remain tightly linked to
task distribution and resource allocation \cite{de2012multiagent,redhead2019dynamics}.
Such recurring patterns of structured coordination, observed across natural and artificial systems \cite{zefferman2023constraints}, are commonly interpreted as manifestations of hierarchy.
Yet how hierarchy arises from initially homogeneous populations, and whether small trait variations can trigger abrupt transitions toward structured inequality, remains an open question. Against this backdrop, we examine how decentralised coordination and reinforcement dynamics, together with a minimal set of ingredients, can generate stable hierarchical organisation in complex adaptive systems.

In this context, hierarchy is not understood as a fixed social ranking or institutional ordering, but as a structural property emerging from interaction patterns.
More specifically, we focus on the formation of persistent directional asymmetries in information and influence flow among agents, arising endogenously through repeated local interactions.

From a modelling perspective, although many theoretical approaches examine how hierarchical structures can arise from initially equivalent individuals \cite{boehm2009hierarchy,redhead2022social,tibbetts2022establishment,haertel2023pathways},
comparatively few provide quantitative frameworks that capture both the stochastic onset of differentiation and the subsequent consolidation of advantage within a unified evolutionary setting, with most existing approaches emphasising either ecological or social mechanisms in isolation.

An important exception is the work of Perret and colleagues, who demonstrate that interactions between individual heterogeneity and group size can shape the emergence of dominance hierarchies in patch-based environments \cite{perret2017emergence,perret2020hierarchy}.
Their framework provides a useful conceptual basis for the present study.

Building on this line of work, we introduce an agent-based modelling (ABM) framework \cite{sayama2015introduction} for a biologically inspired, decentralised multi-agent system.

Within this framework, agents repeatedly interact to express internal capabilities and negotiate transient leadership roles \cite{dridi2018learning,ren2024enhancing,foss2023ecosystem}.
The model integrates ecological, social, and evolutionary processes—combining density-dependent reproduction, stochastic mortality, speaker--listener coordination, and probabilistic resource allocation—to reproduce feedback loops underlying hierarchical organisation.

More broadly, the concept of hierarchy varies across disciplines.
In complex adaptive systems, it often denotes nested, multi-level subsystems that organise information and control flows \cite{lane2006hierarchy,daniel2023multi},
while socio-economic studies emphasise statistical regularities such as rank--size scaling \cite{cristelli2012there,de2021dynamical}.
Despite these differences, both perspectives emphasise differentiation and asymmetry as central features of complex systems.

Consistent with the interactional perspective adopted here, hierarchy is therefore treated functionally as the \emph{directionality of information flow} among agents.
A system is more hierarchical when influence propagates predominantly from higher to lower levels, and less so when interactions are largely reciprocal.
This interpretation emphasises asymmetric organisation rather than fixed positional ranking.

To quantify this interactional notion of hierarchy, we employ the \emph{Trophic Incoherence} (TI) framework \cite{johnson2014trophic,pilgrim2020organisational,rodgers2022network},
a continuous measure of directional order in networks originally developed for ecological food webs.
Low TI values correspond to more ordered, hierarchical structures, whereas higher values indicate flatter or more disordered interaction patterns.
TI therefore provides a principled diagnostic for identifying the onset and persistence of hierarchy in decentralised populations.

Within this framework, we identify the minimal conditions under which hierarchical order can arise and stabilise by systematically varying initial heterogeneity and mutation amplitude.
Adopting Trophic Incoherence as a structural measure further enables the temporal tracking of hierarchy emergence without presupposing fixed leadership or ranking mechanisms.

\section{Methodology and Experimental Design of the Self-Organising Cooperation System}

Our agent-based model (ABM)~\cite{sayama2015introduction} examines the emergence of stable hierarchies from local interactions within a decentralised population, using a minimal environment in which simple behavioural rules give rise to structured, group-level outcomes. Each agent maintains internal states—energy, spatial position, behavioral mode, and an $\alpha$-value representing social standing—and may act independently or in coordination with others.
Agents move, regulate energy, and seek partners; when cooperating, they perform collective foraging and reproduction~\cite{grimaldo2008madem,migliano2022origins}, subject to proximity, energy sufficiency, and behavioral compatibility.
Although cooperation is common, collective activities also introduce competition: access to resources shapes reproductive success, while occupying a leadership role yields greater influence and returns.
Together, these interactions generate selection pressures that couple survival and reproduction with both individual traits and social behaviour.
As a consequence, agents with higher $\alpha$-values more frequently assume the speaker role, which in turn reinforces their standing and forms the basis of our experimental setup for examining the emergence of hierarchical structure in a decentralised multi-agent system.

\subsection{Self-Organisation Principles}

Beyond individual behaviour, the system incorporates ecological feedbacks that regulate population-level stability, with reproduction and survival jointly responding to population density and resource availability~\cite{lomnicki1999individual,gotelli2008primer}.
Population growth follows a standard density-dependent mechanism: as the number of agents $N_t$ approaches the carrying capacity $K$, reproduction slows due to resource limitation and increases again at lower densities. 
The birth rate at time $t$ is defined as
\begin{equation}
    BR_t = f \cdot \left(1 - \frac{N_t}{K}\right),
\end{equation}
where $f \in (0,1)$ denotes the intrinsic fertility parameter, influenced by agent attributes such as age or oestrus status. 
This formulation imposes a smooth upper bound on growth, ensuring a gradual decline in reproduction near environmental capacity.

Mortality follows a Poisson-based process~\cite{lomnicki1999individual}, with risk increasing for older or critically energy-depleted agents. 
Together, these feedbacks create a competitive yet stable ecological setting in which survival and reproduction depend on both internal agent states and overall population density. 
In the baseline configuration, death events occur stochastically, with higher probability for agents exceeding an age threshold ($\tau_t^i \geq 10$) or with zero energy ($E_t^i = 0$), independent of hierarchical position.

\subsection{State Representation and Behavior Dynamics}

In the Multi-Agent System (MAS), each agent interacts with both its environment and other agents through a state-driven process involving sensing, decision-making, and action execution.

\begin{figure}[!t]
\centering
\includegraphics[width=0.73\textwidth]{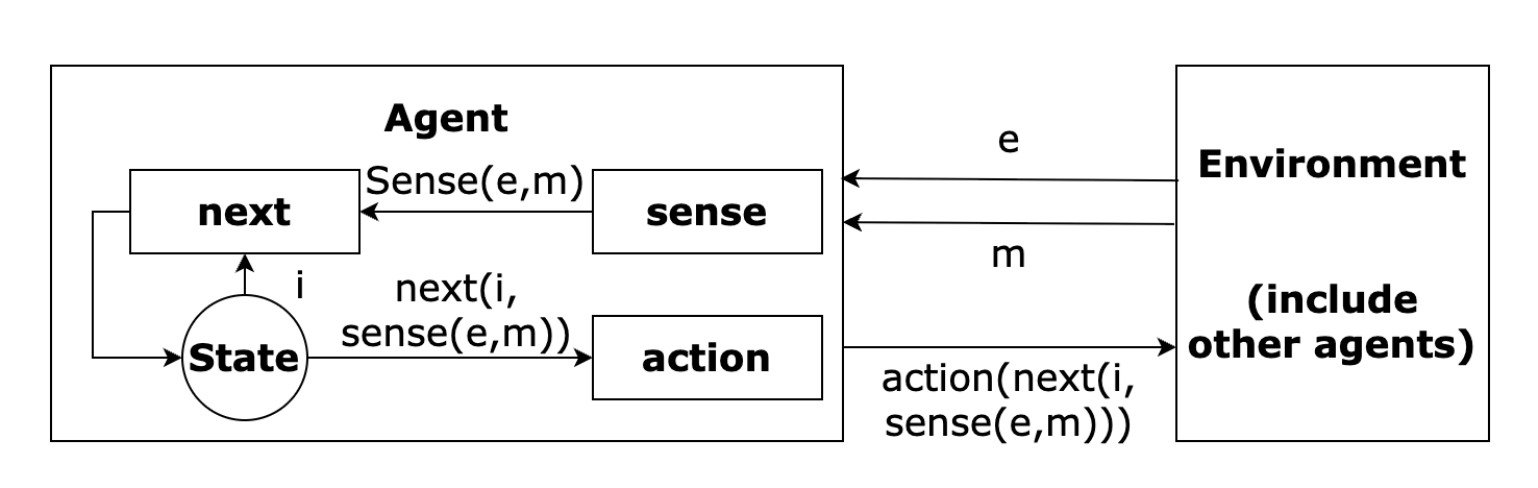} 
\caption{\small 
Agent–environment interaction diagram adapted from \cite{macal2005tutorial,papasimeon2009modelling}, illustrating the information flow between agent states, sensing, and action execution within the MAS. 
}
\label{fig1}
\end{figure}

Fig.~\ref{fig1} illustrates the overall architecture of agent–environment interactions. 
To formally represent these interactions, we define the state of the system at each discrete time~$t$ as a vector~$\underline{S_t}$ that aggregates the states of all agents in the set $\mathcal{N}$:

\begin{equation}
\underline{S_t} = \bigl( S_t^1, S_t^2, \dots, S_t^{\mathcal{N}} \bigr),
\end{equation}
where $S_t^i$ represents the state of agent $i$ at time $t$, and $\mathcal{N}$ is the total number of agents in the system. This formulation provides a complete mapping of the system configuration at each discrete time interval. Here, the superscript indicates the agent index, while the subscript denotes the time index throughout the subsequent equations.

The environment evolves in response to agent actions and system states. In our model, the environment includes spatial features such as resource locations (food), reproduction zones (villages), and agent proximity fields, all of which influence agents’ decisions. Formally, the environmental state at time $t+1$ is given by the environmental response function $T_E(\cdot)$:

\begin{equation}
\underline{\mathcal{E}_{t+1}} = T_E(\underline{S_t}, \underline{\mathcal{E}_{t}}, \underline{A_t}),
\end{equation}
where $\underline{\mathcal{E}_{t}}$ denotes relevant environmental states at time $t$, and $\underline{A_t}$ denotes the actions performed by agents at time $t$. This function captures how collective actions and current conditions jointly shape environmental dynamics. Similarly, the state transition for each agent is determined by the current agent states, environmental conditions, and actions undertaken by agents. The agent state transition function $T_S(\cdot)$ is defined as:

\begin{equation}
\underline{S_{t+1}} = T_S(\underline{S_t}, \underline{\mathcal{E}_t}, \underline{A_t}).
\end{equation}

The transition captures how agent–environment interactions update individual states over time. In the proposed Multi-Agent System (MAS), each agent determines its next action at discrete time step \( t \) through an agent action function \( G(\cdot) \):

\begin{equation}
A_{t+1}^i = G(\underline{S_t}, \underline{\mathcal{E}_t}, \underline{A_t}),
\end{equation}

where \( A_{t+1}^i \) denotes the action of agent \( i \) at time \( t+1 \). The function \( G(\cdot) \) maps current system states, environmental conditions, and past actions to future agent behaviors.

The nature of \( G(\cdot) \) determines whether agent behavior is stochastic or deterministic. All agent actions are updated synchronously at each discrete time step. If \( G(\cdot) \) incorporates probabilistic elements—such as stochastic decision rules or sampling from softmax distributions—the resulting actions \( A_{t+1}^i \) exhibit stochastic characteristics. Similarly, when the environmental update function \( T_E(\cdot) \) includes random or uncertain components, both the environmental state \( \mathcal{E}_{t+1} \) and subsequent agent states \( S_{t+1} \) will also evolve stochastically.

When both the agent function $G(\cdot)$ and the environmental function $T_E(\cdot)$ operate deterministically, the system's evolution becomes fully predictable. 
This structure makes it possible to compare stochastic and deterministic variants of the model under the same update rules. Each agent at time \( t \) maintains a set of internal state variables, including energy level $E_t^i \in \mathbb{R}_{\ge 0}$, sensed environmental information, proximity to resources and settlements, and the availability of neighboring agents for cooperative behavior. These variables collectively guide the agent’s decision-making process and give rise to a range of behavioral modes, which can be categorized into individual and joint actions.

\begin{itemize}
\item \textbf{Single-Agent Actions}: These actions do not require querying the states or previous actions of other agents. They include:
\begin{itemize}
    \item \textbf{DEAD}: Triggered if \( E_t^i = 0 \) or the agent’s age \( \tau_t^i \geq 10 \), at which point the risk of death increases with age. The age of agent $i$ is represented by $\tau_t^i \in \mathbb{N}$, measured in discrete time steps.
    \item \textbf{MOVING\_TO\_VILLAGE}: Occurs when an agent has sufficient energy \((E_t^i > 10)\) but is located beyond a threshold distance \(R_{\text{village}}\) from a village, which could be represented as \( Dv_t^i \geq R_{\text{village}}^m \) . $Dv_t^i$ denotes the Euclidean distance from agent $i$ to the boundary of village $m$, with $m \in \{1,2,3,4,5\}$ and $R_{\text{village}}^m$ the corresponding village radius.
.
    \item \textbf{IDLE}: Default, lowest-priority action when no other conditions are met.
    \item\textbf{WAIT\_REPRO}: if the agent's energy is more than 10 \((E_t^i > 10)\) and its position is inside of one village \( Dv_t^i \leq R_{\text{village}}^m \).
\end{itemize}

\item \textbf{Joint Actions}: These actions involve explicit interactions with and queries of the states or previous actions of other agents:

\begin{itemize}
    \item \textbf{REPRODUCTION  (2 agents)}: Two agents collaborate to reproduce, contingent upon energy levels, proximity to a village, and synchronized previous action states. Agent's sex is encoded as a binary variable, $\text{sex}^i \in \{\text{male}, \text{female}\}$. Mathematically, reproduction occurs only if:

\begin{equation}
\text{REPRO}_{t+1}^{(i,j)} = 
\begin{cases}
1, & \begin{aligned}
     & \text{if } E_t^i > 10 \text{ and } E_t^j > 10 \\
     & \text{and } Dv_t^i \leq R_{\text{village}}^m \\
     & \text{and } Dv_t^j \leq R_{\text{village}}^m  \\
     & \text{and } A_t^i = A_t^j = \text{WAIT\_REPRO} \\
     & \text{and } \text{sex}^i \neq \text{sex}^j
     \end{aligned} \\
0, & \text{otherwise}.
\end{cases}
\end{equation}

    Reproductive success further depends on agents’ relative $\alpha$-scores, influencing offspring number and reinforcing hierarchical dynamics.

    \item \textbf{COOPERATION (3 agents)}: Three agents collaboratively perform tasks like food collection tasks. Initiating such tasks requires agents to evaluate their energy level, nearby idle agents, and available food resources. The collective task initiation criterion is mathematically expressed as:\\
    \begin{equation}
        \text{COOP}_{t+1}^{(i,j,k)} = \\
\begin{cases}
1, & \text{if } E_t^i < 10 \text{ and } \nu_t^i \geq 2 \\
   & \quad \text{and } \mu_t^i \geq 1, \\
0, & \text{otherwise}.
\end{cases}
    \end{equation}
        
    where \(\nu_t^i\) denotes the number of nearby IDLE agents, which $\nu_t^i \in \mathbb{N}$ typically does not exceed 5, and \(\mu_t^i\) denotes the number of accessible food resources (within radius $\rho$ of the agent). 

\end{itemize}
\end{itemize}

\subsection{Inheritance and Mutation in Reproduction}

Following reproduction, a successful pairing between two agents produces one or more offspring whose capabilities arise from a biologically inspired inheritance–mutation process balancing trait transmission and stochastic variability.

At initialization, each agent’s innate capability $C_i$ is drawn from $\mathcal{N}(100, \text{parameter}_c)$, where $\text{parameter}_c$ controls population heterogeneity. Larger $\text{parameter}_c$ values yield broader capability spreads, while smaller ones produce more uniform populations.

Let $C^{p1}>0$ and $C^{p2}>0$ denote the parental capacities of agents $p1$ and $p2$. The offspring’s inherited capability $C^k$ is a weighted combination of the parental maximum and mean, biased toward the stronger parent to model selective inheritance:
\begin{equation}
C^k = h \cdot \max(C^{p1}, C^{p2}) + (1 - h) \cdot \frac{C^{p1} + C^{p2}}{2},
\end{equation}
where $h \in [0,1]$ is the heritability coefficient, set to $h = 0.9$ to represent strong vertical transmission from the fitter parent with minor contribution from the other. Equation~(8) is bounded by parental values, while stochastic mutation introduced in Eq.~(9) prevents monotonic amplification across generations.

To introduce variability, a mutation mechanism perturbs the inherited value with probability $p_m$ by a zero-mean Gaussian noise term:
\begin{equation}
C^{k'} = C^k + \delta,\quad \text{where } \delta \sim \mathcal{N}(0, \sigma^2),
\end{equation}
where $\sigma$ denotes the mutation scale (\text{parameter}$_u$, referred to as the {\em mutation amplitude} in the Results). Increasing $\text{parameter}_u$ amplifies variance and evolutionary exploration, whereas smaller values maintain conservative inheritance. If no mutation occurs, $C^{k'} = C^k$.

Through this process, mutation continuously reintroduces heritable variation across generations, while reproduction biases transmission toward stronger parents, together supporting the gradual consolidation of asymmetric influence patterns in the population.

A viability constraint ensures capability never drops below a functional minimum:
\begin{equation}
\alpha_{t=0}^k = C^{k''} = \max(C^{k'}, 1),
\end{equation}
where the constant $1$ is a lower bound for viable capability rather than a normalization factor. 
We use $\alpha_{t=0}^k$ instead of $C^k$ to highlight that the $\alpha$ notation represents an agent’s social or functional standing, indexed by time $t$. 
Although in the present baseline model $\alpha_t^i$ remains static—reflecting only inherited genetic capability—the time-dependent form anticipates future extensions where ability may evolve through experience accumulation, learning, or reputation updates.

Because initial $C_i$ values are sampled from a normal distribution centered at 100, inherited capacities $C^{k''}$ preserve the parental scale, maintaining heterogeneity across generations and preventing collapse to unit normalization.

\subsection{Speaker Selection in Cooperation}

During cooperative interactions, the speaker functions as a temporary coordination focal point selected endogenously from the interacting group, rather than as a fixed or guaranteed leader. 
Speaker selection is therefore contingent on local interaction dynamics and reflects transient coordination demands rather than persistent authority.

To select a speaker among cooperating agents, we use a softmax-based stochastic rule weighted by ability scores $\alpha_t^i$. 
The probability that agent $i$ becomes the speaker at time $t$ is
\begin{equation}
P_t^i(S) = \frac{\exp(\beta\,\alpha_t^i)}{\sum_{j=1}^{n}\exp(\beta\,\alpha_t^j)},
\end{equation}
where $n$ denotes the number of cooperating agents. The parameter $\beta > 0$ controls the selection pressure: larger $\beta$ increases the probability that agents with higher $\alpha_t^i$ are selected as speakers. This formulation favours agents with higher $\alpha_t^i$ while preserving stochasticity, so leadership emerges probabilistically rather than deterministically. In comparison, a linear proportional rule~\cite{perret2017emergence},
$P_t^i(S) = \frac{\alpha_t^i}{\sum_{j=1}^{n}\alpha_t^j},$
underestimates small ability gaps and saturates when ability distributions are skewed. 
The softmax function increases sensitivity to minor disparities in $\alpha_t^i$, giving slightly stronger agents a proportionally higher chance to lead. 
Conceptually, this corresponds to a multinomial Logit (discrete choice) model in which selection probabilities scale exponentially with perceived ability. 
As a result, hierarchy formation becomes smoother and continuously responsive to marginal competence differences, as reflected in the gradual rise in leadership likelihood for small $\Delta\alpha_t^i$ observed in our results.

\subsection{Consensus-Based Decision-Making and Resource Allocation}

During cooperative tasks, resource distribution is governed by a decentralized consensus process.
Each agent $i$ possesses an ability score $\alpha_t^i$, representing task-relevant competence or influence.
One agent is stochastically selected as the speaker ($s$), while the remaining agents ($\mathcal{L}$) act as listeners.

The group’s response to the speaker is captured by a dynamic \emph{consensus coefficient} $d_{t,q}$, which regulates alignment or resistance during the $q$-th cooperative event at time~$t$:
\begin{equation}
d_{t,q} = 1 + \frac{1}{|\mathcal{L}|}
          \sum_{l \in \mathcal{L}}
          (\alpha_t^s - \alpha_t^l)(d_t^s - d_t^l),
\end{equation}
where $d_t^s$ and $d_t^l$ denote the expected dominance of the speaker and listener, respectively, derived from each agent’s normalized percentile rank $\mathcal{R}_t^i \in (0,1]$ within the cooperating group.
:
\begin{equation}
d_t^s = 0.5 + 0.5\,\mathcal{R}_t^s, \qquad
d_t^l = 0.5\,\mathcal{R}_t^l.
\end{equation}

Higher ranks yield larger $d_t^i$, strengthening influence in the consensus process.
The asymmetric ranges $[0.5,1]$ for speakers and $[0,0.5]$ for listeners allow both dominance and partial resistance.
When the speaker is substantially stronger, $d_{t,q} > 1$ and consensus favors the speaker; when weaker, $d_{t,q} < 1$, indicating resistance.
For small ability differences, $d_{t,q}$ remains near~1, corresponding to neutral consensus.
Although Eq.~(12) introduces local positive feedback, hierarchical amplification is not guaranteed and remains contingent on stochastic speaker selection and broader evolutionary dynamics.

Given $d_{t,q}$, resources are allocated as:
\begin{equation}
r_t^i =
\frac{1 + d_{t,q}\,\alpha_t^i}
     {\sum_{z \in \mathcal{T}} (1 + d_{t,q}\,\alpha_t^z)}.
\end{equation}
This rule amplifies or suppresses ability differences depending on the group’s collective stance.
In each cooperative cycle, $d_{t,q}$ is updated once and $\mathcal{R}_t^i$ recalculated from the current ability distribution, allowing adaptive evolution of inequality.

\begin{table}[h]
\centering
\caption{Model parameters related to the consensus mechanism.}
\begin{tabular}{@{}lll@{}}
\toprule
Symbol & Description & Typical Role \\
\midrule
$\alpha_t^i$ & Individual ability & dynamic variable \\
$\mathcal{R}_t^i$ & Percentile rank within team & (0,1] \\
$d_t^s, d_t^l$ & Expected dominance & speaker/listener \\
$d_{t,q}$ & Consensus coefficient & dynamic \\
$r_t^i$ & Resource share & 0–1 \\
\bottomrule
\end{tabular}
\end{table}

This consensus mechanism links local coordination with emergent inequality and provides a key pathway through which hierarchy stabilizes in the evolving population.

\subsection{Monitoring Hierarchical Emergence via Trophic Analysis}

To quantify the emergence and stability of hierarchical structure in the multi-agent system, we employ the \textit{Trophic Incoherence} (TI) metric. 
Originally developed for food-web analysis, TI has been extended to characterize directed social and interaction networks~\cite{johnson2014trophic,klaise2016trophic,rodgers2022network,pilgrim2020organisational}. 
It measures how much directional links deviate from a strictly ordered hierarchy, providing a continuous index of structural coherence. 
Preliminary ablation checks further confirm that both mutation amplitude and initial heterogeneity are essential—disabling either prevents the system from maintaining low TI, highlighting their indispensable role in sustaining hierarchical order.

Let $W = [w_{ij}]$ denote the weighted adjacency matrix of listener-to-speaker interactions, where each element $w_{ij}$ counts how many times agent $i$ (as a \textbf{listener}) has chosen agent $j$ (as a \textbf{speaker}). 
Each cooperative event thus represents a recommendation, so $W$ encodes accumulated endorsements from follower to leader. 
Since influence flows from speaker to listener, we use the transpose $W^T$ in all calculations.

For each agent $i$, we define:
\begin{equation}
u_i = \sum_j (w_{ij} + w_{ji}), \quad
v_i = \sum_j (w_{ij} - w_{ji}),
\end{equation}
where $u_i$ is the total interaction strength and $v_i$ the net bias between outgoing and incoming links. 
We then construct a symmetric Laplacian-like matrix
$\Lambda = \text{diag}(u_i) - (W + W^T),$
and solve   $h = \Lambda^{\dagger} \cdot v,$
where $h = [h_i]$ gives each agent’s trophic level, $\Lambda^{\dagger}$ is the Moore–Penrose pseudoinverse, and $v = [v_i]$ is the interaction vector.

Trophic incoherence is then:
\begin{equation}
F(h) = \frac{1}{\sum_{i,j} w_{ij}} \sum_{i,j} w_{ij} (h_j - h_i - 1)^2.
\end{equation}

Here, \(F(h) \in [0, \infty)\) quantifies deviation from perfect hierarchy, where \(h_j = h_i + 1\) indicates ideal downstream flow from speaker \(j\) to listener \(i\). 
Lower \(F(h)\) reflects more layered, directional organization; higher values imply flatter or cyclic structures. 
In particular, \(F(h) \to 0\) denotes a perfectly stratified hierarchy, whereas \(F(h) \ge 1\) marks increasing incoherence where directionality is lost. 
This continuous measure tracks how local interactions sustain or disrupt emergent hierarchy over time. 

Self-loops are excluded; for disconnected networks, TI is computed on the largest connected component, with isolated nodes omitted from the \(F(h)\) normalization.
In the following, we refer to \(F(h)\) as TI.

\section{Results and Analysis}

We first examine how the final Trophic Incoherence (TI) varies across combinations of initial heterogeneity ($c$) and mutation amplitude ($u$).
Lower TI values correspond to more stratified, hierarchical states, whereas higher values indicate more disordered configurations.
Figure~\ref{fig:mean_TI} reports the mean final TI after 30\,000 simulation steps, averaged over 20 independent runs, a horizon sufficient for convergence while mitigating stochastic fluctuations.
As $u$ increases, TI decreases markedly, indicating that stronger mutation facilitates the emergence of hierarchical structure.
By contrast, variation in $c$ has a weaker but non-negligible effect, most apparent at low $u$, suggesting that initial heterogeneity primarily influences early formation rather than long-term persistence.

\subsection{Parameter Space Overview}

\begin{figure}[!t]
  \centering
  \includegraphics[width=0.65\linewidth]{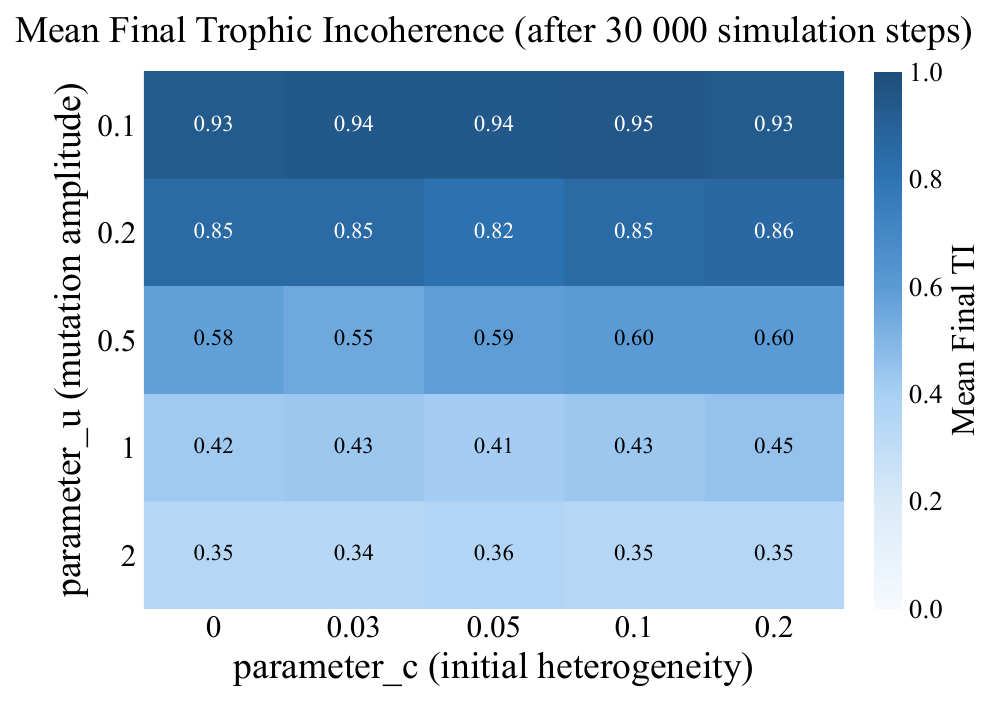}
  \caption{\small
  \textbf{Mean Final Trophic Incoherence across parameter space.}
  Each cell shows the mean final TI across 20 replications.
  Darker blue tones denote lower TI, indicating stronger hierarchical order.
  }
  \label{fig:mean_TI}
\end{figure}

Figure~\ref{fig:mean_TI} summarises the mean final Trophic Incoherence (TI) across the $(c,u)$ parameter space, where lower values indicate more stratified hierarchical organisation and higher values correspond to disordered or egalitarian states.
TI is computed after 30\,000 simulation steps and averaged over 20 independent runs.
As mutation amplitude $u$ increases, TI decreases sharply, indicating that stronger mutation promotes the emergence of hierarchical structure.
By contrast, when mutation is minimal ($u \le 0.1$), TI remains consistently high ($>0.9$) across all $c$ values, and no stable hierarchy forms.
This regime effectively serves as an implicit ablation: in the near absence of stochastic variation, directional differentiation fails to persist, highlighting mutation as a necessary condition for hierarchy formation.

\begin{figure}[!t]
  \centering
  \includegraphics[width=0.65\linewidth]{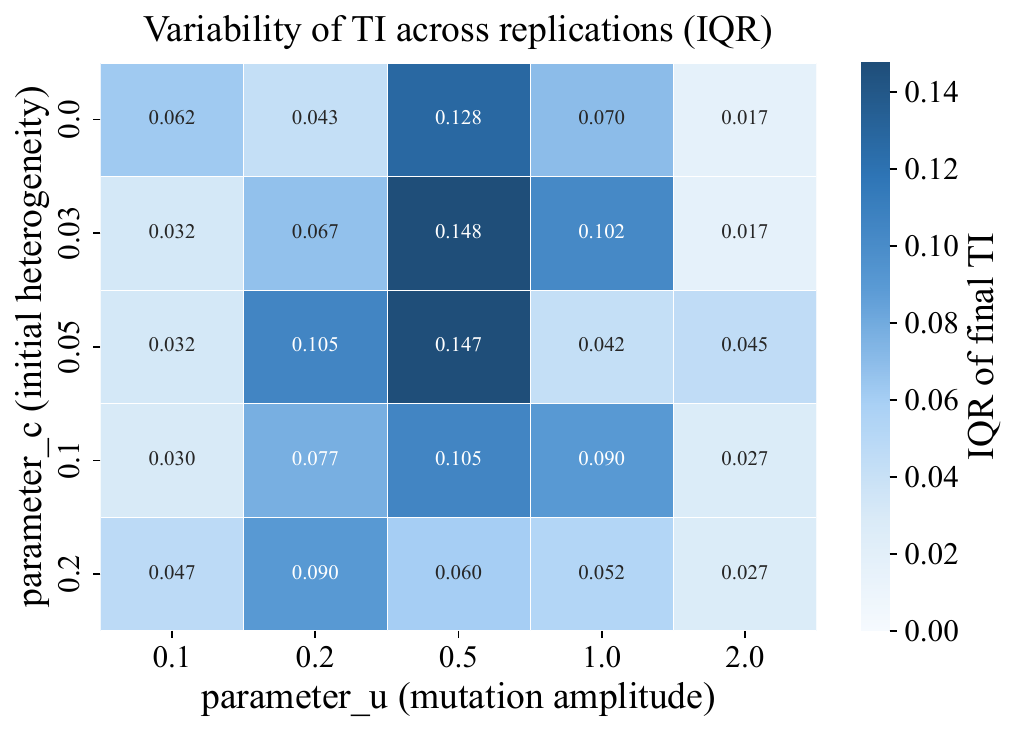}
  \caption{\small
  \textbf{Variability of TI across replications (IQR).}
  The interquartile range captures cross-run variability for each $(c,u)$ pair.
  Smaller values indicate greater stability and convergence of hierarchical order.
  }
  \label{fig:IQR_TI}
\end{figure}

Within the explored parameter range, variation in $c$ exerts a secondary but non-negligible influence—most visible at low $u$—suggesting that initial heterogeneity mainly affects the early onset rather than the long-term persistence of hierarchy.
To assess robustness, we next quantify cross-run variability in the final TI. 
Figure~\ref{fig:IQR_TI} reports the interquartile range (IQR) for each $(c,u)$ pair. 
Smaller IQR values indicate higher stability and convergence of hierarchy formation.

Large mutation amplitudes ($u = 1$ or $u = 2$) produce both low mean TI (Fig.~\ref{fig:mean_TI}) and small IQR, implying that hierarchies formed under these conditions are not only stronger but also highly reproducible. 
In contrast, intermediate mutation levels ($u \approx 0.5$) show the highest variability, revealing a transition zone where outcomes fluctuate between ordered and disordered regimes.
Here, reproducibility refers to \emph{cross-run consistency rather than temporal invariance}, with $\mathrm{IQR}(TI) < 0.05$ indicating consistent outcomes across independent runs.

\begin{figure}[!ht]
  \centering
  \includegraphics[width=0.65\linewidth]{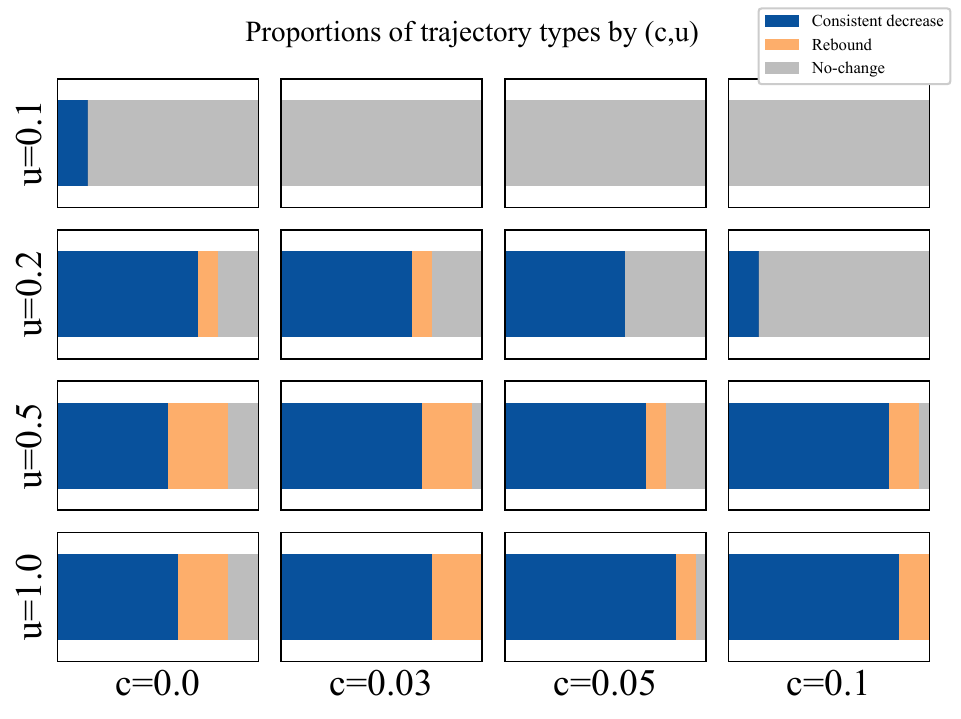}
  \caption{\small
\textbf{Phase map of hierarchical regimes.}
Regions where $\text{median}(TI) < 0.45$ and $\text{IQR} < 0.05$
are classified as Consistent decrease (dark blue),
indicating strong cross-run reproducibility across independent runs;
transitional regions as Rebound (orange),
and disordered regions as No-change (gray).
The map highlights a clear boundary separating ordered and disordered regimes.
}
  \label{fig:phase_map}
\end{figure}

Combining these two indicators, we identify distinct dynamical regimes in the $(c,u)$ space. Figure~\ref{fig:phase_map} shows how these regimes vary across 20 independent runs. As mutation amplitude increases, cross-run consistent trajectories become dominant, 
while rebound and no-change outcomes diminish. 
At intermediate $u$ values ($\approx 0.5$), all three regimes coexist, marking a probabilistic transition between ordered and disordered behavior. 
Together, these three maps (Figs.~\ref{fig:mean_TI}–\ref{fig:phase_map}) reveal a phase-like transition across the parameter space: 
low mutation maintains disordered, high-TI states; intermediate $u$ leads to mixed regimes; and high $u$ consistently produces low-TI, stable hierarchical order.

\subsection{Temporal Dynamics of Hierarchy}

To examine not only \emph{whether} hierarchy emerges but also \emph{how} it develops, we trace the time series of Trophic Incoherence (TI) over the whole simulation.

\begin{figure}[!h]
  \centering
  \includegraphics[width=0.60\linewidth]{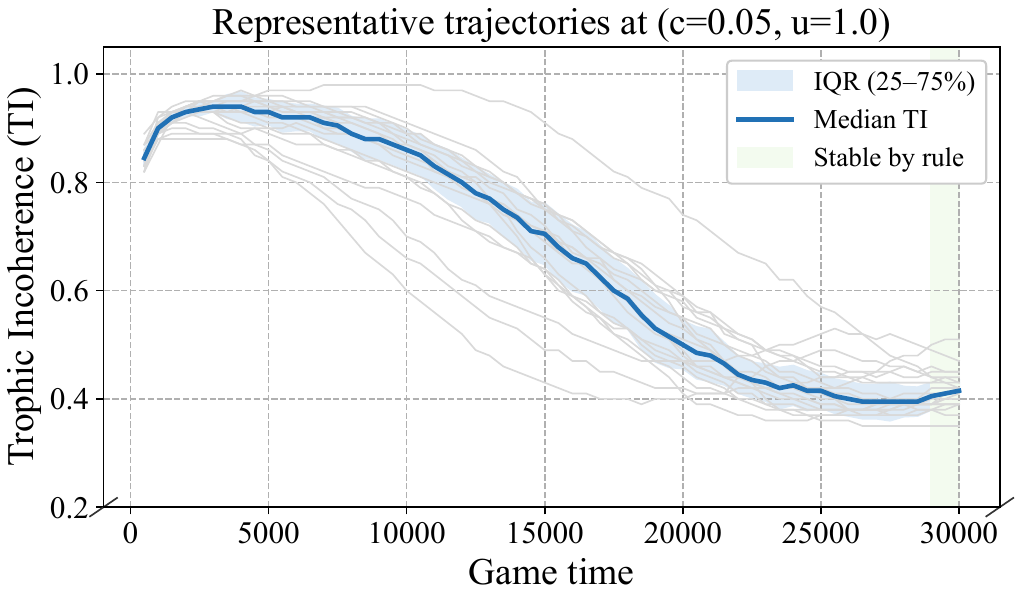}
  \caption{\small
  \textbf{Representative trajectories at $(c=0.05,\,u=1.0)$.}
  Each gray line represents one of 20 runs; the blue curve is the median TI,
  and the shaded area the interquartile range (25--75\%).
  The green region highlights periods satisfying the stability criterion
  ($\text{median}(TI)<0.45$ and $\text{IQR}<0.05$), indicating the onset of a
  stable hierarchical regime.
  }
  \label{fig:rep_traj}
\end{figure}

Figure~\ref{fig:rep_traj} reports a representative case $(c = 0.05,\, u = 1.0)$. At the beginning, TI rises quickly to a high level as interaction links are still noisy and weakly ordered. After this transient, TI starts to decrease monotonically and eventually fluctuates within a narrow band. The blue curve shows that, on average, the system converges toward a low-TI configuration, while the gray trajectories reveal that individual runs may reach this regime at slightly different times. The green shaded segment marks the period in which both criteria $\text{median}(TI) < 0.45$ and $\text{IQR} < 0.05$ are satisfied, i.e. when a stable hierarchical regime has effectively formed.

\begin{figure}[!b]
  \centering
  \includegraphics[width=0.67\linewidth]{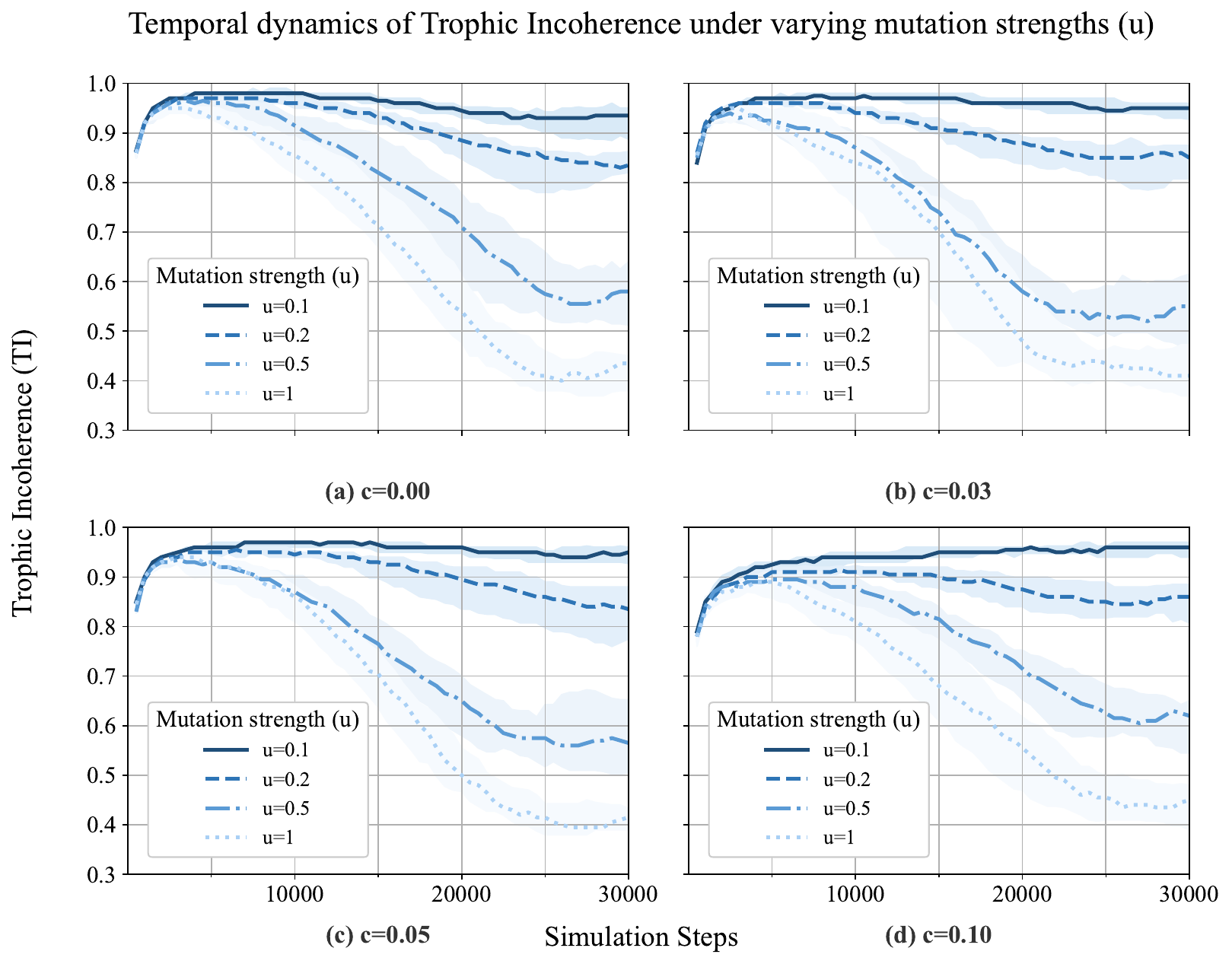}
  \caption{\small
  \textbf{Temporal dynamics of TI under varying mutation strengths $u$.}
  Each curve shows the median TI across 20 replicates (shaded area = IQR 25--75\%);
  rows correspond to different initial heterogeneity $c$.
  Higher $u$ values lead to earlier and deeper ordering, whereas low $u$
  maintains high-TI, weakly ordered states.
  }
  \label{fig:ti-dynamics}
\end{figure}

To see how robust this pattern is across parameter settings, we further compare TI trajectories under different mutation amplitudes $u$ and initial heterogeneity $c$ (Fig.~\ref{fig:ti-dynamics}). A common structure appears in all panels: an early exploration phase with high TI, followed by an ordering phase in which TI declines, and finally a late stage in which TI either stabilizes or rebounds slightly. However, the \emph{speed} and \emph{depth} of this decline depend on $u$. When mutation is weak ($u=0.1$), TI remains high for the whole run, showing that almost no durable hierarchy can be sustained. For intermediate values ($u=0.2$ or $u=0.5$), TI decreases but the spread between runs widens, producing the mixed and “transitional” regimes already identified in Fig.~\ref{fig:phase_map}. When mutation is strong ($u=1.0$), TI drops earliest and reaches the lowest plateau, indicating that continuous trait variation actually helps the system select and reinforce a directional interaction structure. Across different $c$, this temporal profile is largely preserved, confirming that ongoing mutation is a stronger driver of hierarchical ordering than the initial spread of abilities.

In combination with the parameter-space results, these temporal curves show that hierarchy in this model is not created instantaneously but is gradually consolidated as successful speaker--listener relations accumulate in the recommendation network. Mutation that is too weak does not provide enough variation for such relations to be filtered; mutation at moderate to high levels accelerates this consolidation and keeps the system in the low-TI band.

\subsection{Structural Interpretation through Network Comparison}

To illustrate the structural meaning of Trophic Incoherence (TI), we visualize two representative communication networks obtained at the end of the simulation. 
Each network is arranged using a \textit{trophic-level layout} that (i) places lower-level agents near the bottom and (ii) minimizes edge crossings. 
When hierarchy is well defined, these criteria yield clear stratified layers; when it is weak, crossings cannot be minimized without breaking vertical order, producing a tangled and symmetric pattern. 
Thus, the visual contrast between the two panels directly reflects the presence or absence of hierarchical organization rather than mere node rearrangement.

\begin{figure}[!b]
  \centering
  \begin{minipage}[b]{0.43\textwidth}
    \includegraphics[width=\linewidth]{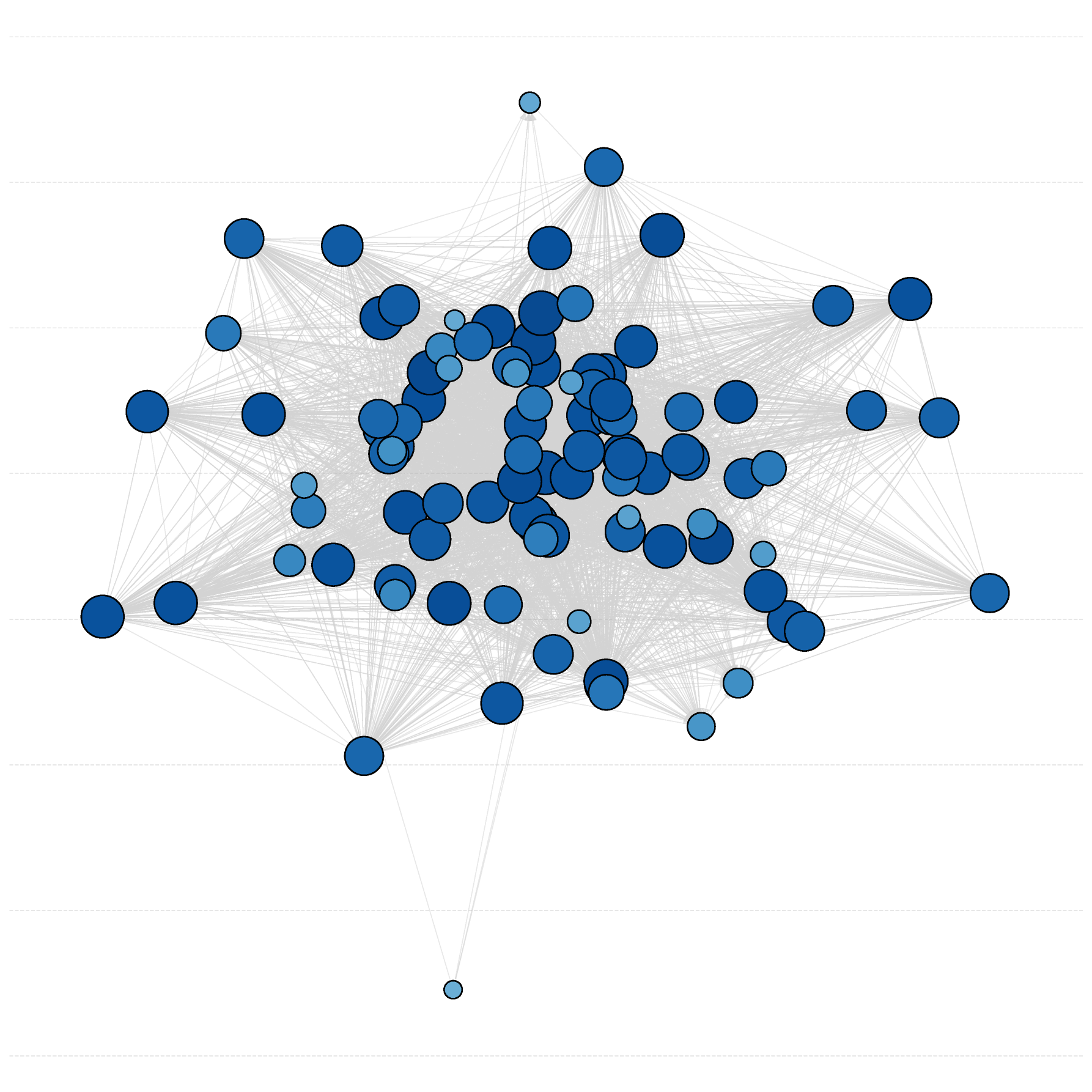}
  \end{minipage}
  \hfill
  \begin{minipage}[b]{0.43\textwidth}
    \includegraphics[width=\linewidth]{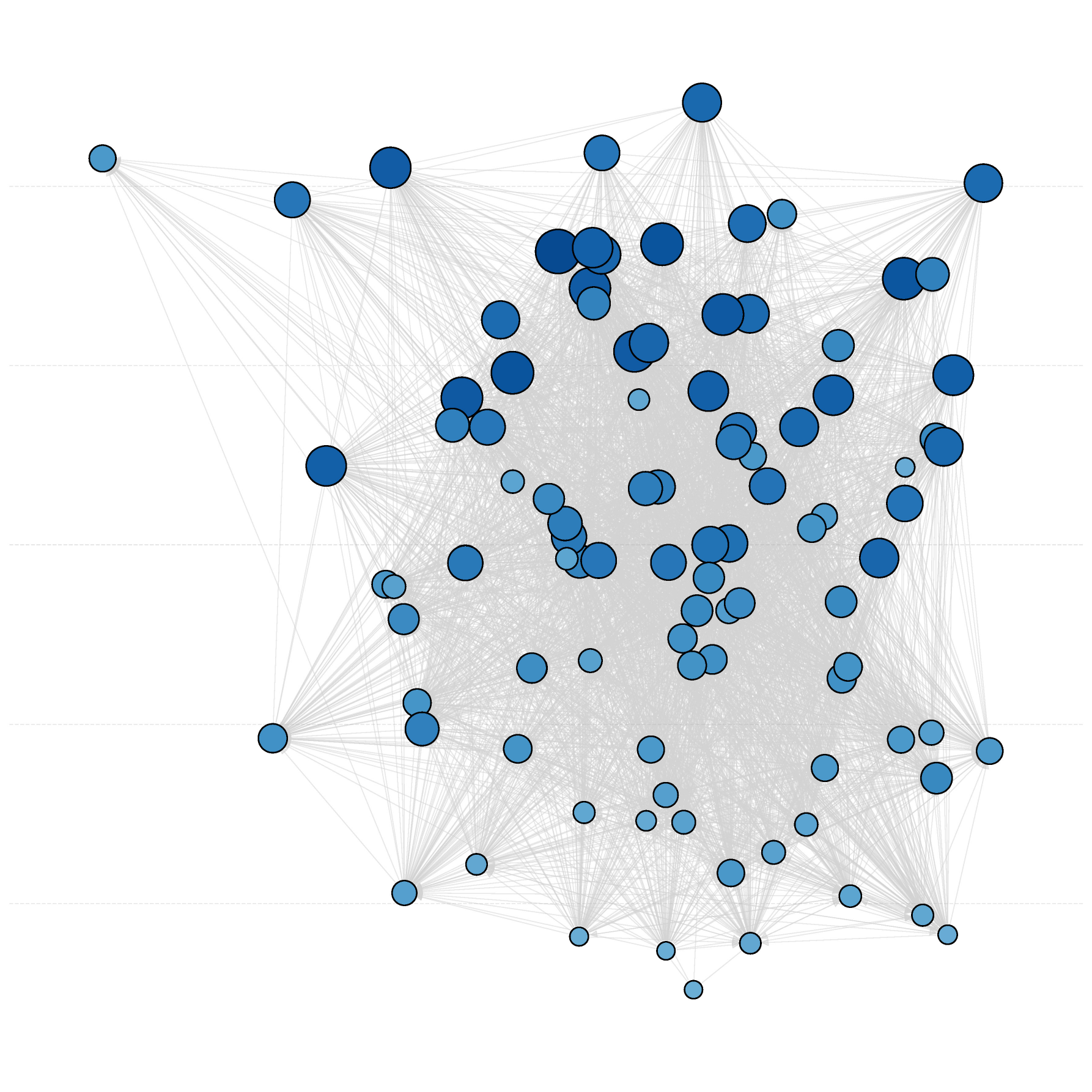}
  \end{minipage}
  \caption{\small
  \textbf{Network Topologies at High and Low TI.}
  (a) Communication network at step 30\,000 with $F(h)\!\approx\!0.99$; 
  (b) at step 30\,000 with $F(h)\!\approx\!0.31$. 
  Both are visualized using a trophic-level layout positioning lower-level nodes at the bottom while minimizing edge crossings. 
  Strong hierarchical ordering, as in (b), forms distinct tiers, whereas incoherent networks like (a) appear tangled and symmetric.
  }
  \label{fig7}
\end{figure}

Each node in Fig.~\ref{fig7} represents a surviving agent positioned vertically by its trophic level \( h_i \). 
Directed edges indicate accumulated listener-to-speaker recommendations, while node size and color encode speaking frequency. 
Larger, darker nodes thus denote agents with higher social endorsement, integrating influence and hierarchical position within the network.

The right-hand network ($F(h)\!\approx\!0.31$) shows clear trophic layering: agents cluster into distinct tiers, and a few dominant nodes receive disproportionate endorsement, reflecting persistent leadership and asymmetric information flow. 
By contrast, the left-hand network ($F(h)\!\approx\!0.99$) appears diffuse and symmetric, lacking both tiers and concentration of influence—an egalitarian configuration without sustained dominance. 
These contrasts illustrate that low TI coincides with visually stratified networks, whereas high TI is associated with diffuse, weakly ordered structures.

\section{Discussion and Future Work}

This study investigated how hierarchical structure can arise in a decentralised multi-agent system without predefined leaders. In our model, agents coordinate through local foraging and speaker--listener interactions, and these interactions gradually induce asymmetric influence patterns captured by the Trophic Incoherence (TI) index.

Across the parameter space, mutation amplitude $u$ plays the dominant role in determining whether hierarchy forms. When mutation is weak, inherited traits remain nearly static and the population stays in a disordered, high-TI regime. As $u$ increases, heritable variation is continuously reintroduced, enabling directional relationships to be filtered and reinforced across generations. This produces a clear transition around $u \approx 0.5$--1.0, where TI drops and stable hierarchy becomes the most likely outcome. Initial heterogeneity $c$ mainly affects the early phase of convergence: higher $c$ delays ordering but does not prevent low-TI states once reinforcement accumulates.

The temporal profiles of TI clarify this mechanism. Under moderate $u$, TI declines steadily as cooperative interactions repeatedly endorse particular speaker--listener pairs, while network visualisations show that low-TI runs develop clear trophic layering and concentrated influence, whereas high-TI runs remain diffuse and symmetric. TI therefore provides a practical link between local coordination dynamics and population-level organisation in the model.

Although each agent’s capability $\alpha^i$ is fixed in the current implementation, its time-indexed notation $\alpha_t^i$ makes the framework compatible with extensions in which ability evolves through learning, reinforcement, or social experience. The present model intentionally prioritises structural hierarchy formation over task-level performance or demographic richness; extensions such as adaptive foraging efficiency, resilience to agent loss, and more diverse gender representations are left for future work.

In summary, hierarchy in this system arises from the interaction of variation, competition, and reinforcement rather than explicit design. Future work will extend the model by introducing experience-based updates to $\alpha_t^i$, allowing environmental conditions to fluctuate, and examining resilience under deception or rapid population turnover. The source code for the simulation model and analysis scripts is publicly available at:
\url{https://github.com/ShanshanMao999/cooperation-to-hierarchy-abm}.

%
%
%

\begin{thebibliography}{10}
\providecommand{\url}[1]{\texttt{#1}}
\providecommand{\urlprefix}{URL }
\providecommand{\doi}[1]{https://doi.org/#1}

\bibitem{boehm2009hierarchy}
Boehm, C.: Hierarchy in the Forest: The Evolution of Egalitarian Behavior. Harvard University Press (2009)

\bibitem{couzin2009collective}
Couzin, I.D.: Collective cognition in animal groups. Trends in Cognitive Sciences  \textbf{13}(1),  36--43 (2009)

\bibitem{cristelli2012there}
Cristelli, M., Batty, M., Pietronero, L.: There is more than a power law in zipf. Scientific reports  \textbf{2}(1), ~812 (2012)

\bibitem{daniel2023multi}
Daniel, P.A., Daniel, E.: Multi-level project organizing: A complex adaptive systems perspective. In: Research handbook on complex project organizing, pp. 138--147. Edward Elgar Publishing (2023)

\bibitem{de2021dynamical}
De~Marzo, G., Gabrielli, A., Zaccaria, A., Pietronero, L.: Dynamical approach to zipf's law. Physical Review Research  \textbf{3}(1),  013084 (2021)

\bibitem{detrain2008collective}
Detrain, C., Deneubourg, J.L.: Collective decision-making and foraging patterns in ants and honeybees. Advances in Insect Physiology  \textbf{35},  123--173 (2008)

\bibitem{dridi2018learning}
Dridi, S., Ak{\c{c}}ay, E.: Learning to cooperate: The evolution of social rewards in repeated interactions. The American Naturalist  \textbf{191}(1),  58--73 (2018)

\bibitem{foss2023ecosystem}
Foss, N.J., Schmidt, J., Teece, D.J.: Ecosystem leadership as a dynamic capability. Long range planning  \textbf{56}(1),  102270 (2023)

\bibitem{gagnon2011high}
Gagnon, S.A., Bruny{\'e}, T.T., Robin, C., Mahoney, C.R., Taylor, H.A.: High and mighty: Implicit associations between space and social status. Frontiers in Psychology  \textbf{2}, ~259 (2011)

\bibitem{gotelli2008primer}
Gotelli, N.J.: A Primer of Ecology. Sinauer Associates (2008)

\bibitem{grimaldo2008madem}
Grimaldo, F., Lozano, M., Barber, F.: Madem: a multi-modal decision making for social mas. In: Proceedings of the 7th international joint conference on Autonomous agents and multiagent systems (AAMAS). vol.~1, pp. 183--190. IFAAMAS (2008)

\bibitem{haertel2023pathways}
Haertel, T.M., Leckelt, M., Grosz, M.P., Kuefner, A.C., Geukes, K., Back, M.D.: Pathways from narcissism to leadership emergence in social groups. European Journal of Personality  \textbf{37}(1),  72--94 (2023)

\bibitem{johnson2014trophic}
Johnson, S., Domínguez-García, V., Donetti, L., Muñoz, M.A.: Trophic coherence determines food-web stability. Proceedings of the National Academy of Sciences  \textbf{111}(50),  17923--17928 (2014)

\bibitem{klaise2016trophic}
Klaise, J., Johnson, S.: From neurons to epidemics: How trophic coherence affects spreading processes. Chaos: An Interdisciplinary Journal of Nonlinear Science  \textbf{26}(6),  065310 (2016)

\bibitem{lane2006hierarchy}
Lane, D.: Hierarchy, complexity, society. In: Hierarchy in natural and social sciences, pp. 81--119. Springer, Dordrecht (2006)

\bibitem{macal2005tutorial}
Macal, C.M., North, M.J.: Tutorial on agent-based modeling and simulation. In: Proceedings of the Winter Simulation Conference, 2005. pp. 14--pp. IEEE (2005)

\bibitem{migliano2022origins}
Migliano, A.B., Vinicius, L.: The origins of human cumulative culture: from the foraging niche to collective intelligence. Philosophical Transactions of the Royal Society B  \textbf{377}(1843),  20200317 (2022)

\bibitem{papasimeon2009modelling}
Papasimeon, M.: Modelling agent-environment interaction in multi-agent simulations with affordances. Citeseer (2009)

\bibitem{perret2020hierarchy}
Perret, C., Hart, E., Powers, S.T.: From disorganized equality to efficient hierarchy: how group size drives the evolution of hierarchy in human societies. Proceedings of the Royal Society B  \textbf{287}(1928),  20200693 (2020)

\bibitem{perret2017emergence}
Perret, C., Powers, S.T., Hart, E.: Emergence of hierarchy from the evolution of individual influence in an agent-based model. In: Artificial Life Conference Proceedings. pp. 348--355. MIT Press (2017)

\bibitem{pilgrim2020organisational}
Pilgrim, C., Guo, W., Johnson, S.: Organisational social influence on directed hierarchical graphs, from tyranny to anarchy. Scientific Reports  \textbf{10}(1), ~4388 (2020)

\bibitem{redhead2019dynamics}
Redhead, D., Cheng, J.T., Driver, C., Foulsham, T., O'Gorman, R.: On the dynamics of social hierarchy: A longitudinal investigation of the rise and fall of prestige, dominance, and social rank in naturalistic task groups. Evolution and Human Behavior  \textbf{40}(2),  222--234 (2019)

\bibitem{redhead2022social}
Redhead, D., Power, E.A.: Social hierarchies and social networks in humans. Philosophical Transactions of the Royal Society B  \textbf{377}(1845),  20200440 (2022)

\bibitem{ren2024enhancing}
Ren, T., Zeng, X.J.: Enhancing cooperation through selective interaction and long-term experiences in multi-agent reinforcement learning. arXiv preprint arXiv:2405.02654  (2024)

\bibitem{rodgers2022network}
Rodgers, N., Ti{\v{n}}o, P., Johnson, S.: Network hierarchy and pattern recovery in directed sparse hopfield networks. Physical Review E  \textbf{105}(6),  064304 (2022)

\bibitem{sayama2015introduction}
Sayama, H.: Introduction to the modeling and analysis of complex systems. Open SUNY Textbooks (2015)

\bibitem{tibbetts2022establishment}
Tibbetts, E.A., Pardo-Sanchez, J., Weise, C.: The establishment and maintenance of dominance hierarchies. Philosophical Transactions of the Royal Society B  \textbf{377}(1845),  20200450 (2022)

\bibitem{de2012multiagent}
de~Weerdt, M.M., Zhang, Y., Klos, T.: Multiagent task allocation in social networks. Autonomous Agents and Multi-Agent Systems  \textbf{25}(1),  46--86 (2012)

\bibitem{zefferman2023constraints}
Zefferman, M.R.: Constraints on cooperation shape hierarchical versus distributed structure in human groups. Scientific Reports  \textbf{13}(1), ~1160 (2023)

\bibitem{lomnicki1999individual}
Łomnicki, A.: Individual-based models and the individual-based approach to population ecology. Ecological Modelling  \textbf{115}(2-3),  191--198 (1999)

\end{thebibliography}
%
\bibliographystyle{splncs04}

\end{document}